\documentclass[12pt,preprint]{aastex}

\slugcomment{Accepted to the Astronomical Journal}

\def\MO{M_\odot}
\def\LO{L_\odot}

\def\HCO{\mathrm{H}^{13}\mathrm{CO}^{+}}

\def\MD{\dot{M}}

\shorttitle{ALMA view of G0.253+0.016}
\shortauthors{A.E.Higuchi et al.}

\begin{document}
\title{ALMA view of G0.253+0.016: \\
Can cloud-cloud collision form the cloud?}

\author{Aya E. Higuchi\altaffilmark{1,2}, James O. Chibueze\altaffilmark{2,3}, 
Asao Habe\altaffilmark{4}, Ken Takahira\altaffilmark{4}, Shuro Takano\altaffilmark{5}}
\email{aya.higuchi@nao.ac.jp}

\altaffiltext{1}{Joint ALMA Observatory, Alonso de C{\'o}rdova 3107, Vitacura, Santiago, Chile}
\altaffiltext{2}{National Astronomical Observatory of Japan 2-21-1 Osawa, Mitaka, Tokyo, 181-8588, Japan}
\altaffiltext{3}{Department of Physics and Astronomy, Faculty of Physical Sciences, University of Nigeria, Carver Building, 1 University Road, Nsukka, Nigeria}
\altaffiltext{4}{Department of Physics, Faculty of Science, Hokkaido University, Kita 10 Nishi 8 Kita-ku, Sapporo 060-0810, Japan}
\altaffiltext{5}{Nobeyama Radio Observatory, Nobeyama, Minamimaki, Minamisaku, Nagano 384-1305, Japan}

\begin{abstract}

We present the results of sulfur monoxide, SO line emission observations of G0.253+0.016 with the Atacama Large Millimeter/submillimeter Array (ALMA) at an angular resolution of 1\arcsec.7. The dense and massive molecular cloud of G0.253+0.016 is highly sub-structured, yet shows no obvious signs of cluster formation. We found three outstanding features of the cloud from the SO emission, namely, shell structure of radius 1.3 pc, large velocity gradients of 20~km~s$^{-1}$~pc$^{-1}$ with the cloud, and cores with large velocity dispersions (30--40~km~s$^{-1}$) around the shell structure. We suggest that these large-velocity dispersion cores will form high-mass stars in the future. In attempt to explore the formation scenario of the dense cloud, we compared our results with numerical simulations, thus, we propose that G0.253+0.016 may have formed due to a cloud-cloud collision process.

\end{abstract}


\section{INTRODUCTION}

Most stars, particularly high-mass stars ($>$$\,$8$~\MO$) form in stellar clusters (Lada $\&$ Lada 2003). 
Stellar clusters form in dense and massive molecular clumps (size$\sim$$\,$1~pc, mass$\sim$$\,$100--1000~$\MO$, density
$\sim$10$^{3-5}$~$\rm{cm}^{-3}$) (Ridge et al. 2003; Lada $\&$ Lada 2003; Higuchi et al. 2009; Higuchi et al. 2010; Higuchi et al. 2013). 
Dense gas in the cluster forming clumps are significantly dispersed by the feedback from newly formed stars, thus blurring our understanding of cluster formation.
To explore the initial conditions and details of cluster formation, it is necessary to study molecular clouds in the early stages of cluster formation.

One of the unresolved issues of cluster formation is how dense clouds evolve into the compact and massive stellar clusters like the ``Arches" cluster 
(Lis $\&$ Menten 1998).
The formation mechanism of such a super-star cluster is mysterious because they are expected to form in massive and small clouds of 1~pc scale. 
Therefore, without an external triggering process, it may be impossible to form super-star cluster.
In recent literatures,  cloud-cloud collisions are argued to be responsible for super-star cluster formation 
(Furukawa et al. 2009, Fukui et al. 2009, Ohama et al. 2010). 
Furukawa et al. (2009) suggested that young cluster, Westerlund 2, is likely to form by cloud-cloud collision mechanism.
Westerlund 2 is a remarkable galactic super-star cluster, containing more than 10 massive stars with total stellar mass of 4500 $\MO$ 
in a small volume of only 1~pc in radius. The age of Westerlund 2 is suggested to be few Myrs (Ascenso et al. 2007; Furukawa et al. 2009; Rauw et al. 2007). 

G0.253+0.016 (hereafter G0.25) is very massive ($\sim$2$\times$10$^{5}\MO$, radius of 2.8~pc; Longmore et al. 2012) and dense 
($\sim$3$\times$10$^{4}$~cm$^{-3}$; Kauffmann et al. 2013), 
with low dust temperature of 23~K (Rodr{\'{\i}}guez $\&$ Zapata 2013) and no obvious signs of cluster formation.
G0.25 forms part of 100 pc circum-nuclear ring of clouds (Molinari et al. 2011) at 8.5$\,$kpc distance (Longmore et al. 2012).
G0.25 presents an interesting site for the study of the initial condition of super-star cluster formation, because it is more massive and dense than the Orion A cloud, but hardly forms stars at all (Lis et al. 1994). 
The infrared luminosity of the entire cloud is 3$\times$10$^{5}$$\LO$, suggesting the presence of at least 5 embedded stars at evolutionary stages earlier than B0 (Lis et al. 2001).
Rodr{\'{\i}}guez $\&$ Zapata (2013) also suggested that there are no O stars associated with the cloud.
Recently, Kauffmann et al. (2013) presented N$_{2}$H$^{+}$ results arguing that G0.25 is presently far from forming high-mass stars and clusters.

In this paper, we present 1\farcs7 resolution image of SO(v=0 3(2)--2(1)) line emission from the ALMA Cycle 0 observations. 
SO molecular line emissions are known to trace relatively dense and active clouds in a region 
(Bachiller et al. 2001; Aladro et al. 2013; Hacer et al. 2013; Codella et al. 2014). 
We investigated the spatial distributions and velocity structures with high-resolution observations based on ALMA archive data of G0.25 obtained with an extended configuration. The mosaic of the molecular line emission across this cloud were obtained at 90 GHz. We speculate that G0.25 cloud may represent the precursor to a super-star cluster.

\section{OBSERVATIONS}

G0.25 was observed with ALMA (Hills et al. 2010) during Early Science Cycle 0 with extended configuration, using the Band 3 receivers. 
The array was in a configuration with projected baselines length between $\sim$ 36.1 to $\sim$ 452.8 m, 
sensitive to maximum angular scales of 9\farcs1 and providing a synthesized beam of 2\farcs30 $\times$ 1\farcs53. 
The field of view was $\sim$ 65\farcs5. 
A total of six data sets were collected, using 24 antennas of 12$\,$m diameter and accounting for 4 hours of total integration time on target.
Weather conditions were good and stable, the system temperature varied from 50 to 60 K.

The correlator was set up to four spectral windows in dual polarization mode, 
centered at 89.10000~GHz, 87.20000~GHz, 99.10000~GHz, and 101.10000~GHz. 
The effective bandwidths used was 1875~MHz, for each of the spectral windows, 
corresponding to velocity resolutions of 3.286, 3.357, 2.954, and 2.896 km s$^{-1}$, respectively, after Hanning smoothing. 
In this work we present results from the observations in the SO(3(2)--2(1)): 99.29987~GHz emission line.

The ALMA calibration includes simultaneous observations of the 183 GHz water line with water vapor radiometers 
that measure the water column in the antenna beam, later used to reduce the atmospheric phase noise. 
Amplitude calibration was done using Neptune, and quasars J1717-337 and J2225-049, and NRAO$\,$530 were used to calibrate 
the bandpass and the complex gain fluctuations, respectively. Data reduction was performed using the version 3.4 of the 
Common Astronomy Software Applications (CASA) package (http://casa.nrao.edu), 
and visibilities imaged and deconvolved with 3.4 km s$^{-1}$ channels. 
We used the task CLEAN to image the visibilities. 
The cubes were imaged with a uniform 3.4 km s$^{-1}$ spectral resolution for an rms noise in line-free channels of 
1.1 to 1.4 mJy beam$^{-1}$ per 3.4 km s$^{-1}$ channel.

\section{RESULTS}
\subsection{Spatial distributions of the SO emission}

\begin{figure}
\epsscale{0.9}
\plotone{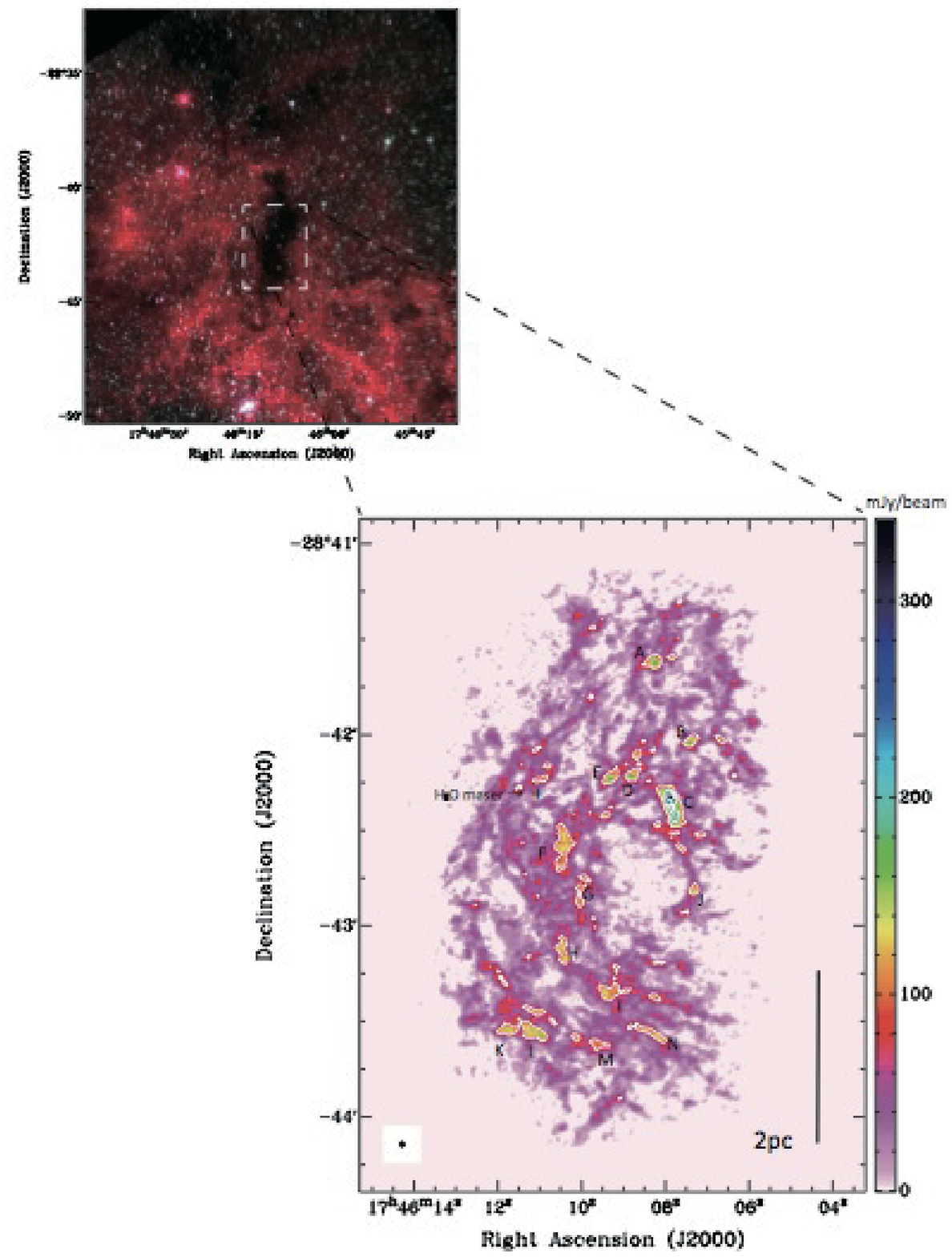}
\caption{
Top: Image of the $\it{WISE}$ 3.4$\,$$\micron$, 4.6$\,$$\micron$, and 22$\,$$\micron$ composite color image.
Bottom: Integrated intensity map of the SO(v=0 3(2)--2(1)) emission in a range of $-$60 to 110 km$\,$s$^{-1}$ (color and contours).
The contours with the intervals of the 30 $\sigma$ levels start from 
the 30 $\sigma$ levels, where the 1 $\sigma$ noise level is 3 mJy/beam.
The H$_{2}$O maser reported by Lis et al. (1994) is marked.
The black line shows 2 pc scale. 
The filled circle at the bottom left corner shows the effective resolution of 1\arcsec.7.} 
\label{map1}
\end{figure}

Figure \ref{map1} shows the integrated intensity map of the SO(v=0 3(2)--2(1)) emission.
There is a clear shell-like structure of radius of $\sim$ 1.3 pc in the integrated intensity map of SO.
We made the map using only interferometric dataset, which implies that we are only seeing features of the shell-like structure within the size of 9\farcs1. 
All components above this scale were resolved out.
Comparing our SO map with the dust continuum map traced by SCUBA 450~$\micron$ in Longmore et al. (2012), we found the strong dust continuum traces the shell-like structure within the cloud. 
From the comparisons, we interpreted that the shell structure traced by SO emission is not an artifact of the interferomtric observations.
We defined the SO emission enclosed in the 30$\,$$\sigma$ contours of the integrated intensity map as SO cores. 
We named these cores which are effectively larger than the spatial resolution of the observations, as Core~$\it{A}$,~$\it{B}$ to $\it{N}$ 
(see Figure \ref{map1}).
SO emission have been detected in some shocked regions, 
e.g., outflow shocked region in L\,1157 (Bachiller et al. 2001), tracing the cavity opened by the jet in NGC\,1333 (Codella et al. 2014), and 
NGC\,1068 (Aladro et al. 2013), implying that observed shell structure in G0.25 may have been due to shock related activity in the region.
We also found that there are hub-filament structures and cores (core~$\it{C}$ to $\it{I}$, and $\it{N}$), which are connected to the shell-like structure. 
Comparing Figure \ref{map1} with the $\it{WISE}$ image of G0.25, 
we see that the SO emissions trace the network of pc-long filaments seen in dust extinction. 
Thanks to high resolution offered by the ALMA telescope, the image of the network of pc-long filaments have been clarified.

The SO abundance is thought to be enhanced in high-temperature environment and/or in the late stage of molecular cloud evolution.
These characteristics are due to the formation of SO mainly via neutral-neutral reactions such as 
\begin{eqnarray}
\label{so}
\rm{S} + \rm{OH} \rightarrow \rm{SO} + \rm{H}.
\end{eqnarray}
Neutral-neutral reactions are generally slower than ion-molecule reactions and tend to have activation energies, 
which must be overcome during the reactions.
The effects of neutral-neutral reactions to SO formation were reviewed previously in a paper by Takano et al. (1995).
In the above point of view, the regions with enhanced SO intensity in the obtained ALMA image can be in relatively dense,  
late stage of molecular cloud evolution, and/or high-temperature. The non-detection of N$_2$H$^+$ (Kauffmann et al. 2013) 
around the cores $\it{C}$ to $\it{H}$ and $\it{J}$, 
which are SO peak regions by Kauffmann and collaborator could be due to relatively higher temperatures of the shocked region.
The dense regions can accelerate the formation of SO via slow neutral-neutral reactions by increasing the collision rate between the reactants.

\subsection{Velocity structures of the cores traced by SO emission}
\subsubsection{1st and 2nd moment maps}

\begin{figure}
\epsscale{1}
\plotone{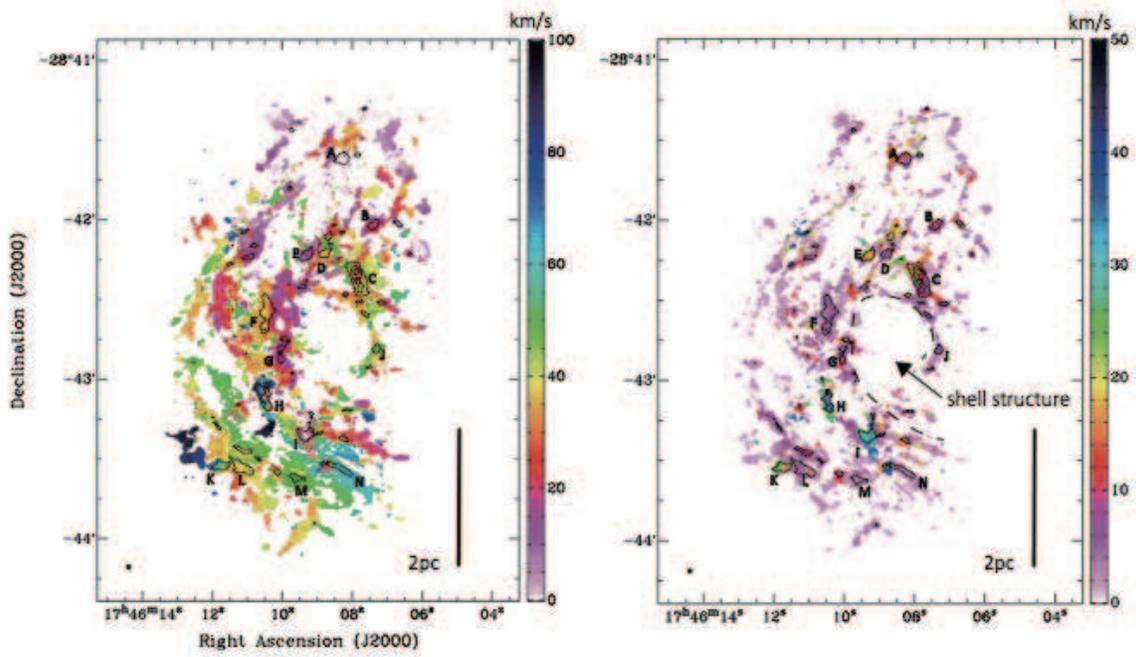}
\caption{Left: Integrated intensity map of the SO(v=0 3(2)--2(1)) emission (contours) superposed on 
the first moment map (color).
Right: Integrated intensity map of the SO(v=0 3(2)--2(1)) emission (contours) superposed on 
the second moment map of the SO(v=0 3(2)--2(1)) emission (color).
The contours with the intervals of the 30 $\sigma$ levels start from 
the 30 $\sigma$ levels, where the 1 $\sigma$ noise level is 3 mJy/beam.
The black line shows 2 pc scale. 
The filled circle at the bottom left corner in each panel shows the effective resolution of 1\arcsec.7.}
\label{map2}
\end{figure}

In order to investigate the detailed velocity structures of the cloud, unveil the presence or absence of velocity gradients, 
identify cores with large velocity dispersion 
($\sigma_{v}$, $\sigma_{v}$ is the one-dimensional velocity dispersion), we produced 1st and 2nd moment maps (Figure \ref{map2}).
From the 1st moment image, cores $\it{A}$ to $\it{F}$ and $\it{I}$ have velocities of 0 to 40 km$\,$s$^{-1}$, 
and cores $\it{K}$ to $\it{N}$ and $\it{H}$ of the cloud have 40 to 100 km$\,$s$^{-1}$.
The overall velocity structure of the cloud has a velocity difference of $\sim$100 km$\,$s$^{-1}$. 
We estimated the velocity gradient ($\sigma_{\rm{grad}}$), $v_{\rm{diff}}$/$2R$, where $v_{\rm{diff}}$ is the velocity difference of $\sim$100 km$\,$s$^{-1}$ and $R$ is the radius of the cloud, to be $\sim$20~km~s$^{-1}$~pc$^{-1}$ within the whole cloud.

Furthermore, from the 2nd moment images, we found the cores associated with SO large velocity dispersions (${\sigma}_{v}$$\sim$~30 to 40~km$\,$s$^{-1}$) 
along the shell-like structure, which may indicate the interactions between molecular gas and external effects.
Within the filament, cores $\it{C}$, $\it{H}$, $\it{I}$, $\it{K}$ and $\it{N}$ have large velocity dispersions. 
The velocity dispersions observed along the shell-structures are most likely due to external shock, 
while the dispersion observed within the filament could be associated with very early phases of star formation considering the presence of $\sim$ 22 GHz H$_2$O masers (Kauffmann et al. 2013).
This is the first time that such large velocity dispersions are observed in G0.25.
The mechanisms of enhancement of large velocity dispersions are unclear.
However there is no evidence of the existence of H {\sc ii} region and external effects (e.g., stellar wind, super nova remnant).

\subsubsection{Position-Velocity diagram}

\begin{figure}
\epsscale{1}
\plotone{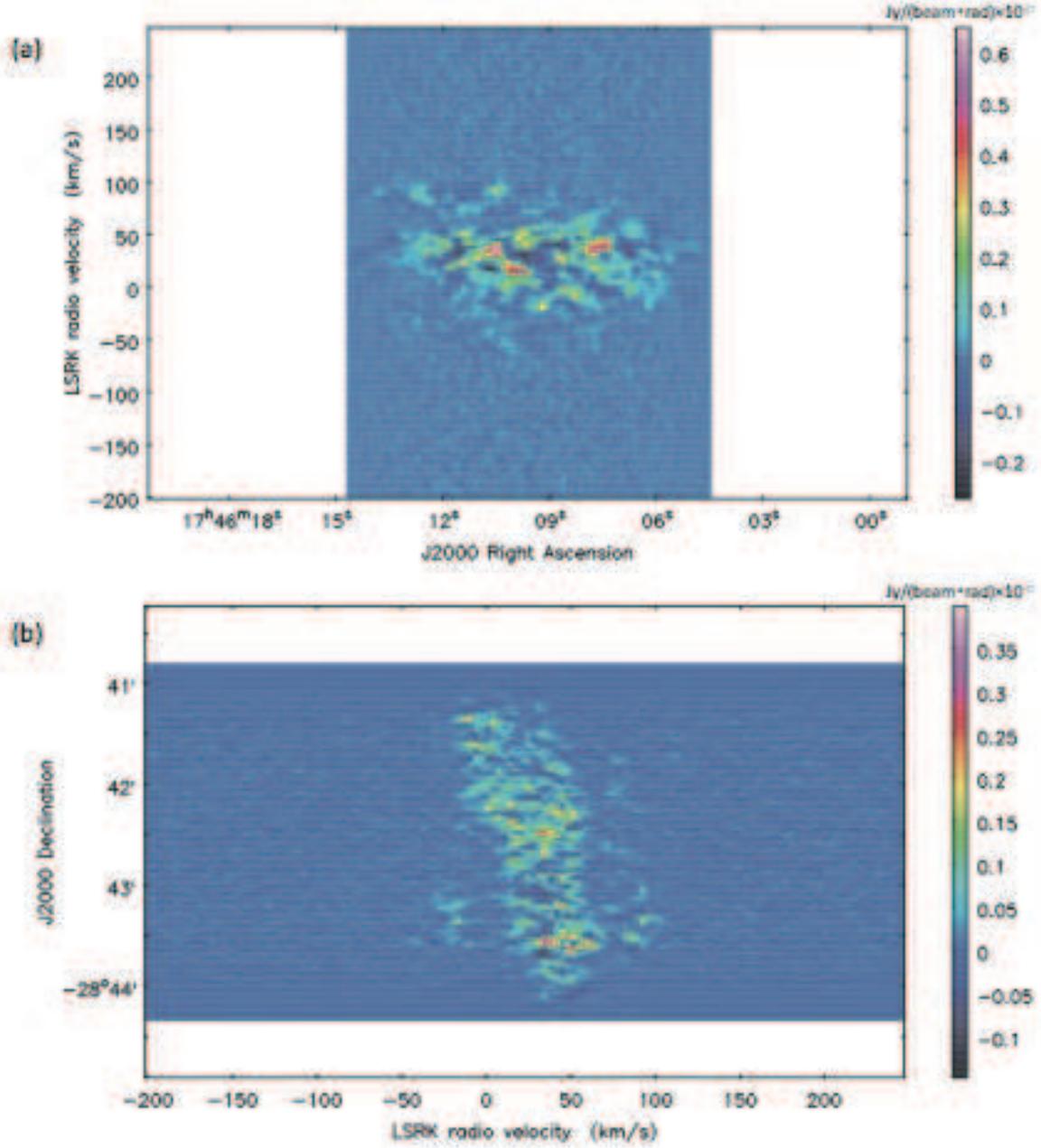}
\caption{Integrated intensity images of SO(v=0 3(2)--2(1)) collapsed along each spatial axis of the cube, (a)
intensity as a function of right ascension (R.A.) and velocity and of (b) declination and velocity.}
\label{pvmap}
\end{figure}

\begin{figure}
\epsscale{1}
\plotone{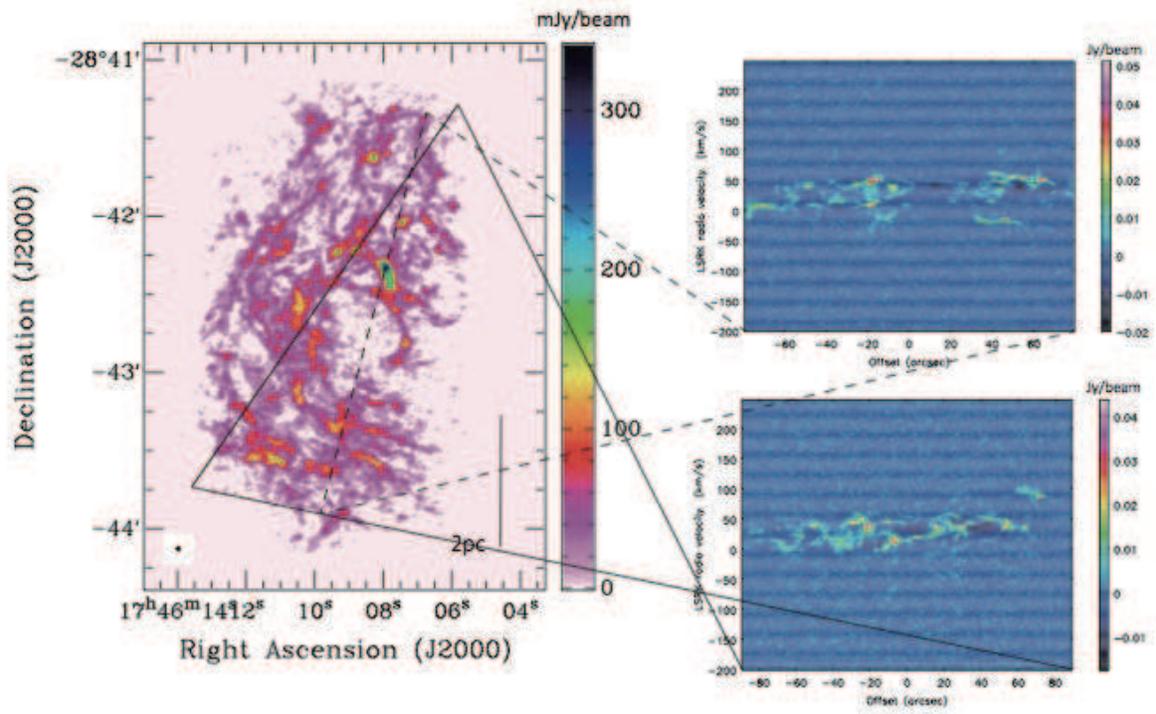}
\caption{
Left: Integrated intensity map of the SO(v=0 3(2)--2(1)) emission.
The right panels are the position-velocity (P-V) diagrams of SO(v=0 3(2)--2(1)) emission.}
\label{pvmap2}
\end{figure}

In order to investigate the velocity structures of the SO cores in detail, we made two position-velocity (P-V) diagrams 
(see Figure \ref{pvmap}).
Figure \ref{pvmap}(a) shows integrated intensity SO emission as a function of both right ascension (R.A.) and velocity,
and (b) as a function of declination and velocity, 
i.e., the data cube collapsed along each spatial axis. 
Figure \ref{pvmap} (a) shows a ``hole" around V$_{LSR}$ $\sim$ 30 km~s$^{-1}$, an indication that of the observed shell-like structures.
In Figure \ref{pvmap}(b), we obtained bulk linear velocity gradients with multiple components. 
These could be multiple components with different velocities on a rigid-like rotation.
We also made P-V diagrams for G0.25 cloud with the different position angles (see Figure \ref{pvmap2}).
The axis of the slice covers the whole SO shell structure. 
From these results, we found that there are multiple components with different velocities at different position angles.

\section{DISCUSSION}\label{4}
\subsection{Kinematic structures of the cloud}

We found that G0.25 cloud have velocity gradients in the 1st moment maps as shown in Figure \ref{map2}.
As discussed in the Section 3.2, we found the following kinematic signatures;
(1) there exists distinct velocity gradients, 
(2) they have multiple velocity components on a rigid-like rotation, and 
(3) large velocity dispersion cores are distributed mainly around the shell structure, with a few such cores within the filaments. 
We regard the above kinematic condition as being one of the key structures in massive cloud formation. 
We examined the energy balance of the cloud using the equilibrium virial theorem 
(e.g., Goodman et al. 1993, Stahler $\&$ Palla 2005).

\begin{eqnarray}
\label{balance}
\it{2K + 2U + W + B = 0},
\end{eqnarray} 
where $\it{K}$ is the kinetic energy, $\it{U}$ is the energy contained in random, thermal motion, 
$\it{W}$ is the gravitational potential energy, and $\it{B}$ is the energy of magnetic field.
Here, we ignore the effect of the magnetic field.
From the description of Stahler $\&$ Palla (2005), we form the ratios, $\it{K}$/$\it{|W|}$ and $\it{U}$/$\it{|W|}$ as below.
\begin{eqnarray}
\label{kin}
\frac{K}{|W|} \sim 0.5\,\left(\frac{\Delta{V}}{4~\rm{km~s}^{-1}}\right)^{2}\left(\frac{M}{10^{5}~\MO}\right)^{-1}\left(\frac{R}{\rm{25~pc}}\right)
	\nonumber \\
	\quad \quad \quad
    = 0.1\,\left(\frac{\sigma_{v}}{4~\rm{km~s}^{-1}}\right)^{2}\left(\frac{M}{2\times10^{5}~\MO}\right)^{-1}\left(\frac{R}{\rm{2.8~pc}}\right),
\end{eqnarray}
where $\Delta{V}$ is the velocity width in FWHM ($\Delta{V}$=$\sqrt{8\ln{2}}\sigma_{v}$), ${M}$ is the mass, and ${R}$ is the radius of the cloud.
We used mean value of the velocity dispersion of SO (${\sigma_{v}}$$\sim$4~km~s$^{-1}$) in the 2nd moment map.
\begin{eqnarray}
\label{therm}
\frac{U}{|W|} \sim 3\times10^{-3}\,\left(\frac{M}{10^{5}~\MO}\right)^{-1}\left(\frac{R}{\rm{25~pc}}\right)\left(\frac{T}{\rm{15~K}}\right)
	\nonumber \\
	\quad \quad \quad
	= 2\times10^{-2}\,\left(\frac{M}{2\times10^{5}~\MO}\right)^{-1}\left(\frac{R}{\rm{2.8~pc}}\right)\left(\frac{T}{\rm{20~K}}\right),
\end{eqnarray}
where ${T}$ is the representative gas temperature of the cloud.
From our estimation, the cloud tends to be in a gravitationally stable. 

Meanwhile, the velocity gradients are seen in the entire G0.25 cloud. 
We compared our result with the previous studies of velocity gradients of clouds in Larson (1983) and Goodmann et al. (1993), 
which shows the plot of $\sigma_{\rm{grad}}$ as a function of the cloud size.

\begin{eqnarray}
\label{grad}
\sigma_{\rm{grad}} = 0.33 \left(\frac{R}{\rm{2.8~pc}}\right)^{-0.4 \pm 0.2} [\rm{km~s}^{-1}~\rm{pc}^{-1}],
\end{eqnarray}
where $\sigma_{\rm{grad}}$ is the velocity gradient of the cloud, $R$ is the radius of the cloud.
If we apply the cloud size of 2.8~pc, the velocity gradient becomes to be 0.33~km~s$^{-1}$~pc$^{-1}$, 
which is an order of magnitude smaller than our result of 20~km~s$^{-1}$~pc$^{-1}$.
Such a large velocity gradient in the cloud is strange compared with the previous studies of the star forming regions.
These observational results imply that these velocity structures are related to the initial conditions in the cloud, 
and must be preserved during the ages (several 10$^{5}$yr: Kauffmann et al. 2013).

\subsection{Comparison with the simulation}

The filamentary structures and cores of SO in G0.25 cloud are very similar to results of cloud cloud collision simulations. 
Habe $\&$ Ohta (1992) simulated head-on collision between non-identical clouds with supersonic relative velocities in order to investigate star formation triggered by cloud-cloud collisions. They assumed that mass ratio of  both clouds is 1:4 and the radius ratio is 1:2. They considered internal random motions of clouds as the effective sound speeds. The clouds collided each other with Mach numbers of the relative speeds, 5$\sim$10. Since their simulation was carried out assuming   the isothermal equation of state, we can change their physical parameters of the small cloud into the radius of $\sim$1.5~pc, and mass of $\sim$0.5$\times10^5 M_{\odot}$, and the larger one into the  radius of $\sim$3~pc, and mass of $\sim$2$\times$10$^{5}$~M$_{\odot}$. Their relative speeds are $\sim$30-60~km~s$^{-1}$, which are consistent with our result. They found that a gravitational instability is triggered by a non-identical cloud-cloud collision. They showed that the large cloud is disrupted and shell structures are formed by the bow shock induced by the collision with the small cloud, and that the small cloud is compressed by the shock induced by the collision with the large cloud and its gravitational collapse is triggered. 
Anathpindika et al. (2010) also discussed the stability of the bow-shock resulting from their simulation of head-on collision between clouds of different sizes. 
Three-dimensional simulations of similar collisions between non-identical clouds were performed by Takahira et al. (2014) submitted that explored formation of filamentary structures and dense cores in the collision between clouds of different masses under varying collision velocities. They also show the formation of the bow shock at the collision interface, the inclusion of turbulence producing a jagged shell of shocked gas. This fragmented into multiple dense cores, which rapidly accreted while in the shocked gas to form objects that could be the precursors to massive stars. These features are very similar to our observation. 
Inoue et al. (2013) demonstrate that massive and gravitationally unstable molecular cloud cores are formed behind the strong shock waves 
induced by cloud-cloud collision. 
The conclusions drawn here may also be relevant to the shells driven by powerful winds from young super star-clusters or expanding supernovae shells. 
However, there are no evidence of such an external effects in G0.25 cloud, we suppose that cloud-cloud collision is a reasonable mechanism to trigger massive cloud formation for G0.25. Figure 5 shows the schematic image of the cloud-cloud collision for G0.25.

Previous studies also proposed the cloud-cloud collision mechanism for G0.25 cloud in the discussion session of Kauffmann et al. (2013). Lis $\&$ Menten (1998) and Lis et al. (2001) take the existence of widespread SiO emission as evidence for an ongoing cloud-cloud collision. The cloud-cloud collision mechanism is previously considered as an efficient mechanism to trigger star formation, but it also leads the destruction of the cluster-forming systems rather than gravitational collapse. However, Furukawa et al. (2009) reported super-star cluster formation triggered by the cloud-cloud collision mechanism. 
From the recent results, cloud-cloud collision is reasonable mechanism to form the initial condition of the super-star cluster, e.g., Infrared Dark Cloud.

\subsection{Implication: existence of large-${\sigma}_{v}$ cores}

Ikeda et al. (2007) carried out $\HCO$(1--0) core survey, 
covering the entire Orion A molecular cloud using the Nobeyama 45 m radio telescope.
They found three cores toward the M42 H {\sc ii} region with velocity dispersions (${\sigma}_{v}$), 
significantly larger than those of the other cores, suggesting that the energy input from the H {\sc ii} region increases the velocity dispersion.
The large-${\sigma}_{v}$ cores have the potential to form the most massive stars because the mass infall rate is proportional 
to the third power of the velocity width ($\MD$=0.975${c_{\rm{eff}}}^{3}$/G: from McKee $\&$ Tan 2003).
From the 2nd moment map of SO, we found that the areas of large velocity dispersion 
(${\sigma}_{v}$$\sim$ 30 to 40 km$\,$s$^{-1}$) are distributed along the shell-like structure.
From the above results, we consider that these regions possibly contain large-${\sigma}_{v}$ cores, which will form high-mass stars.
SO emission have been detected in the active and shocked region (e.g., Bachiller et al. 2001; Aladro et al. 2013; Codella et al. 2014).
We interpreted that Kauffmann et al. (2013) proposed that G0.25 cloud has a lack of dense cores, which will form stars immediately. 
These cores will be detected by dense gas tracers (e.g., $\HCO$, N$_{2}$H$^{+}$, 10$^{5-6}$~cm$^{-3}$; Kauffmann et al. 2013). 
However SO molecule is considered to trace more active cores than N$_{2}$H$^{+}$ cores.
It will take time to form stars because the SO cores will be gravitationally bound due to the turbulence decay.
Thus, we understand that G0.25 cloud is in the very early stage of the star formation, triggered by the cloud-cloud collision.

\section{CONCLUSIONS}

We present the results of sulfur monoxide, SO, line emission observations of G0.25 with the ALMA, 
at an angular resolution of 1\arcsec.7. 
Our results and conclusions are summarized as follows:

\begin{enumerate}
\item We presented the SO map with a size of $\sim$ 3$^{\prime}$$\times$1.5$^{\prime}$ for G0.25 cloud.
We identified detailed filamentary structures and cores within the cloud with high resolution observations.

\item We discovered the shell structure of radius 1.3 pc with the SO emission.
SO emission have been detected in some shocked regions, 
e.g., outflow shocked region in L\,1157 (Bachiller et al. 2001), tracing the cavity opened by the jet in NGC\,1333 (Codella et al. 2014), 
and NGC\,1068 (Aladro et al. 2013), implying 
that observed shell structure in G0.25 may have been due to shock related activity in the region.

\item We found large velocity gradients of $\sim$ 20~km~s$^{-1}$~pc$^{-1}$ within the cloud, 
and cores with large velocity dispersions ($\sim$ 30 -- 40~km~s$^{-1}$) around the shell structure. 
We suggest that these high-velocity dispersion cores will form high-mass stars in the future.

\item
We estimated the virial ratios taking into account the contribution of rotation, 
and found that G0.25 cloud is likely to be gravitationally bound condition.
However large velocity gradient in the G0.25 cloud is strange compared with the previous studies of the star forming regions.
In order to explain the physical conditions of G0.25, we compared our results with numerical simulations, 
we propose that G0.253+0.016 may have formed due to a cloud-cloud collision process.

\end{enumerate}

\bigskip

We thank the referee, Steven Stahler for constructive comments that helped to improve this manuscript. 
We also thank the ALMA staff for these observations during the commissioning stage.
This paper makes use of the following ALMA data: ADS/JAO.ALMA 2011.0.00217.S. 
ALMA is a partnership of ESO (representing its member states), NSF (USA) and NINS (Japan), together with NRC (Canada) and NSC and ASIAA (Taiwan), 
in cooperation with the Republic of Chile. The Joint ALMA Observatory is operated by ESO, AUI/NRAO and NAOJ.
Finally, we acknowledge Elizabeth Tasker and Norikazu Mizuno for their contributions to our study.


\begin{figure}
\epsscale{0.6}
\plotone{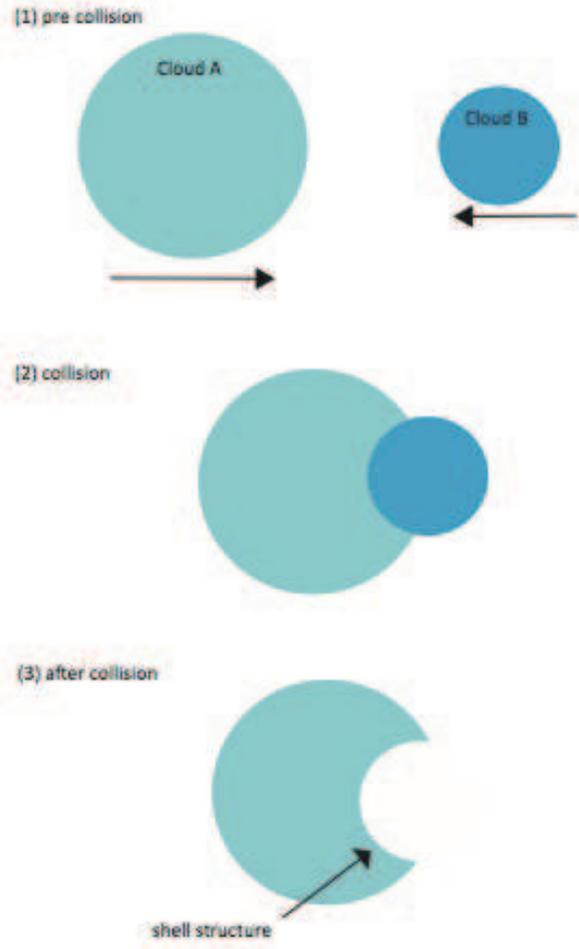}
\caption{Schematic picture of triggering massive cloud formation by cloud-cloud collision.
Small cloud has radius of $\sim$1.5 pc, mass of $\sim$0.5$\times$10$^{5}$$\,$$\MO$, 
and the larger one has radius of $\sim$3 pc, mass of $\sim$2$\times$10$^{5}$$\,$$\MO$.
They collided with the relative speed of $\sim$30-60$\,$km$\,$${\rm{s}}^{-1}$.}
\label{image}
\end{figure}

\end{document}